\documentstyle[epsfig,cite]{article}
 \makeatletter
 \let\chapter\hid@chapter
 \makeatother
 \newcommand{\gevsq  }{\mbox{$\rm GeV^2$}}
 
 \newcommand{\almz   }{\mbox{$\alpha_{s}(M^{2}_z)$}}
 \newcommand{\ftp    }{\mbox{$F_{\mathrm{2}}^{\mathrm{p}}$}}
 \newcommand{\ft     }{\mbox{$F_{2}^{\gamma}$}}
 \newcommand{\fl     }{\mbox{$F_{\rm L}^{\gamma}$}}
 \newcommand{\fkqed  }{\mbox{$F_{\mathrm{k,QED}}^{\gamma}$}}
 \newcommand{\faqed  }{\mbox{$F_{\mathrm{A,QED}}^{\gamma}$}}
 \newcommand{\fbqed  }{\mbox{$F_{\mathrm{B,QED}}^{\gamma}$}}
 \newcommand{\ftxq   }{\mbox{$\ft(x,\qsq)$}}
 \newcommand{\flxq   }{\mbox{$\fl(x,\qsq)$}}
 \newcommand{\ftqed  }{\mbox{$F_{\mathrm{2,QED}}^{\gamma}$}}
 \newcommand{\epem   }{\mbox{$\rm e^+e^-$}}
 \newcommand{\qsq    }{\mbox{$Q^{2}$}}
 \newcommand{\psq    }{\mbox{$P^{2}$}}
 \newcommand{\aem    }{\mbox{$\alpha$}}
 \newcommand{\aemsq  }{\mbox{$\aem^2$}}
 \newcommand{\mumu   }{\mbox{$\mu^+\mu^-$}}
 \newcommand{\az     }{\mbox{$\chi$}}
 \newcommand{\qqbar  }{\mbox{$\mathrm{q\bar{q}}$}}
 \newcommand{\lsim   }{\raisebox{-0.5mm}{$\stackrel{<}{\scriptstyle{\sim}}$}}
 \begin{document}
 \title{\bf
        The photon structure function measurements from LEP
       }
 \author{Richard Nisius, on behalf of the LEP 
         Collaborations~\footnote{Invited 
         talk given at the International Europhysics
         Conference on High Energy Physics, 19-26 August 1997, Jerusalem, 
         Israel, to appear in the Proceedings.}\\
         PPE Division, CERN, CH-1211 Gen{\`e}ve 23, Switzerland}
 \maketitle
 \vspace{-7cm}
 \begin{flushright} {\mbox{ }} \end{flushright}
 \vspace{+5cm}
\begin{abstract}
 The present knowledge of the structure of the photon 
 based on measurements of photon structure functions is 
 discussed.
 This review covers QED structure functions and the hadronic structure 
 function \ft.
\end{abstract}
%
%
\section{Introduction}
\label{sec:intro}
 One of the most powerful tools to investigate the internal structure 
 of quasi real photons is the measurement of photon structure functions
 in deep inelastic electron photon scattering at \epem\ colliders. 
 These measurements have by now a tradition of sixteen years
 since the first \ft\ was obtained by PLUTO~\cite{PLU-8102}. 
 The LEP accelerator is a unique place for the measurements of 
 photon structure functions until a high energy linear collider is realised.
 It is unique because of the large coverage in \qsq\ owing to the various beam 
 energies covered within the LEP programme, and 
 due to the high luminosities delivered to the experiments.
 Structure function measurements are performed by all four LEP experiments, 
 however concentrating on different aspects of the photon structure.
 The main idea is that by measuring the differential cross section 
 \begin{equation}
  \frac{d^2\sigma_{e\gamma\rightarrow e X}}{dxdQ^2}
 =\frac{2\pi\aemsq}{x\,Q^{4}}
  \left[\left( 1+(1-y)^2\right) \ftxq - y^{2} \flxq\right]
 \end{equation}
 one obtains the photon structure function \ft\ which 
 is proportional to the parton content of the photon and therefore
 reveals the internal structure of the photon.
 Here \qsq\ is the absolute value of the four momentum squared of the 
 virtual photon, $x$ and $y$ are the usual dimensionless variables of 
 deep inelastic scattering and \aem\ is the fine structure constant.
 In the region of small $y$ studied ($y\ll 1$) the contribution 
 of the term proportional to \flxq\ is small and it is usually neglected.
 \par
 Because the energy of the quasi-real photon is not known, 
 $x$ has to be derived by measuring the invariant mass of the 
 final state $X$, which consists of \mumu\ pairs for \fkqed, $k=2,A,B$,
 and of hadrons created by a \qqbar\ pair in studies of \ft.
%
 The invariant mass can be determined
 accurately in the case of \mumu\ final states, 
 and measurements of \fkqed\ are statistically limited.
 For hadronic final states the measurement of $x$ 
 is a source of significant uncertainties
 which makes measurements of \ft\ mainly systematics limited.
%
%
\section{The QED structure functions} 
\label{sec:lepton}

%
 \begin{figure}[htb]
 \begin{center}
 \epsfig{figure=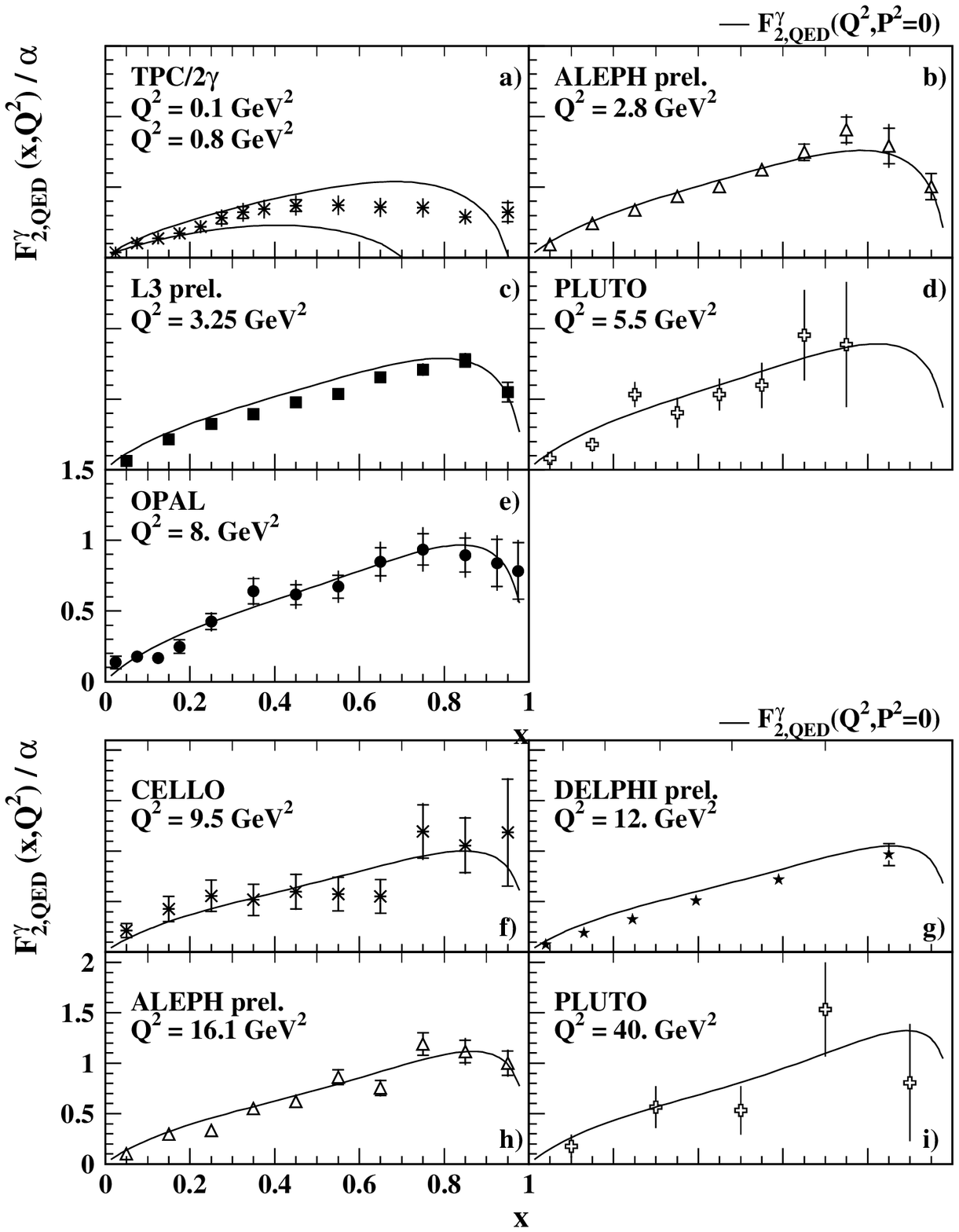,width=12.cm}
 \caption{%
 \it{
 The measurements of \ftqed\ as a function of $x$ for various \qsq\
 compared to QED assuming $\psq=0$. The \qsq\ 
 values for the predictions are taken from the publications.
 If the data were unfolded to a given \qsq\ this value is taken,
 if only the average \qsq\ of the sample is given, c), this value 
 is taken, and if no value at all is quoted the \qsq\ range as obtained 
 from the information of the event selection is shown, a).
 The quoted errors for g) are statistical only.
 \label{fig:fig1}
 }
 }
 \end{center}
 \end{figure}
%
 Several measurements of QED structure functions from LEP have been 
 submitted to this conference.
 The \mumu\ final state is such a clean environment that it allows for much 
 more subtle measurements to be performed than in the 
 case of hadronic final states. 
 Therefore the 
 investigation
 of QED structure functions serves not only as a test of 
 QED but rather it is used to refine the experimentalists tools in a real
 but clean environment to investigate the possibilities of extracting 
 similar information from the much more complex hadronic final states.
 \par
 \az\ is defined as the azimuthal angle between the plane 
 spanned by the muon pair and the plane spanned by the incoming quasi-real 
 photon and the deep inelastically scattered electron.
 This angle has been used to extract structure functions of the photon,
 \faqed\ and \fbqed, which can not be assessed by the cross section 
 measurement, which is dominated by the contribution of \ftqed.
 Azimuthal correlations based on \az\ 
 were investigated by ALEPH (prel.)~\cite{ALE-EGM01}, 
 L3 (prel.)~\cite{L3C-JER01} and OPAL~\cite{OPALPR182}, 
 and \ftqed\ has been obtained by all four experiments.
 Figure~\ref{fig:fig1} shows the world summary of 
 the \ftqed\ measurements~\cite{TPC-8401,CEL-8301,PLU-8501,OPALPR088,DEL-9601,%
 ALE-EGM01,L3C-JER01}.
 All measurements are consistent with expectations from QED. The LEP 
 data are so precise that it is possible to study the effect of the small 
 virtuality \psq\ of the quasi real photon~\cite{DEL-9601,L3C-JER01}.
%
%
\section{The hadronic structure function \ft}
\label{sec:hadron}
%
 \begin{figure}[htb]\begin{center}
 \epsfig{figure=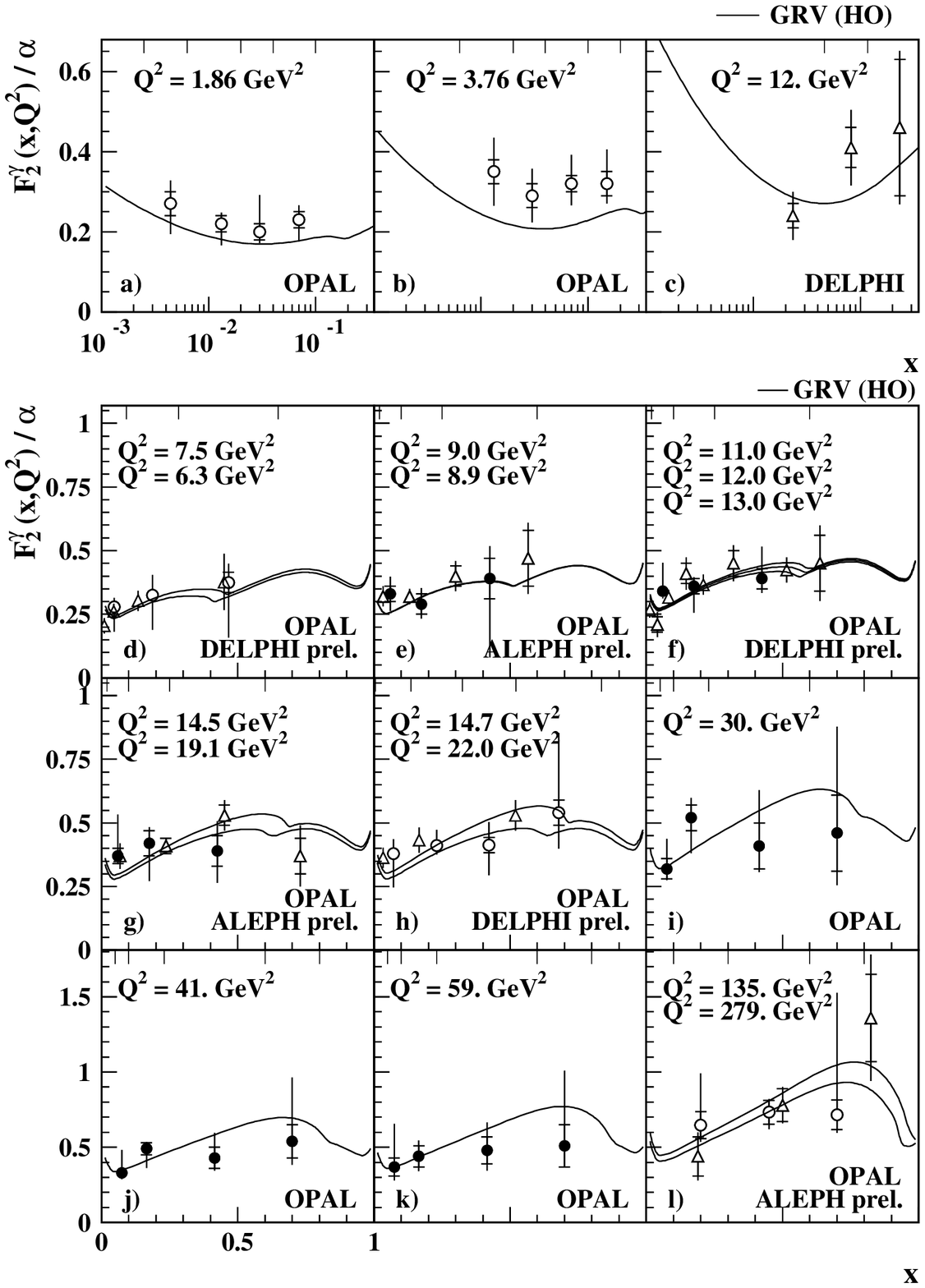,width=12.0cm}
 \caption{%
 \it{
 The measurements of \ft\ as a function of $x$ unfolded on a 
 logarithmic $x$ scale, a)--c), or on a linear $x$ scale, d)--l),
 compared to the prediction of the GRV (HO) parametrisation.
 The OPAL data at 11~\gevsq\ (41~\gevsq) are the combined data 
 from 9 and 14.5~\gevsq\ (30 and 59~\gevsq). 
 The inner error bar is the statistical error and the outer the 
 quadratic sum of statistical and systematic error.
 (ALEPH, DELPHI triangles, OPAL circles)
 \label{fig:fig2}
 }
 }
 \end{center}\end{figure}
%
 The measurement of \ft\ has attracted a lot of interest at LEP over 
 the last years. 
 The LEP Collaborations have measured \ft\ in the range 
 $0.0025<x$ \lsim\ 1 
 and $(1.86<\qsq<279)$ \gevsq, the largest range ever 
 studied~\cite{OPALPR185,OPALPR207,OPALPR213,TIA-9701,ALE-JER01}.
 This work has also encountered some 
 difficulties~\cite{OPALPR185,ALE-JER01} which were not 
 considered in older determinations of \ft.
 Two distinct features of the photon structure are investigated. 
 Firstly the shape of \ft\ is measured as a function of $x$ at fixed \qsq.
 Particular emphasis is put on 
 measuring
 the low-$x$ behaviour of \ft\ in comparison 
 to \ftp\ as obtained at HERA.
 Secondly the evolution of \ft\ with \qsq\ at medium $x$ is investigated.
 This evolution is predicted by QCD to be logarithmic.
 In general \ft\ is found to rise smoothly
 towards large $x$ and there is some weak indication for a possible rise 
 at low $x$ for $\qsq <4$~\gevsq, as shown in Fig~\ref{fig:fig2}.
 This behaviour is satisfactorily described (except in Fig~\ref{fig:fig2}b)
 by several of the existing \ft\ 
 parametrisation, e.g. GRV~\cite{GLU-9201,GLU-9202} and SaS~\cite{SCH-9501}.
 Experimentally there seems to emerge an inconsistency between the 
 OPAL~\cite{OPALPR213} and PLUTO~\cite{PLU-8401} data on one hand
 and the TPC~\cite{TPC-8701} data on the other hand at 
 $\qsq \approx 4$~\gevsq, see~\cite{OPALPR213}.
 \par
\begin{figure}[htb] \begin{center}
 \epsfig{figure=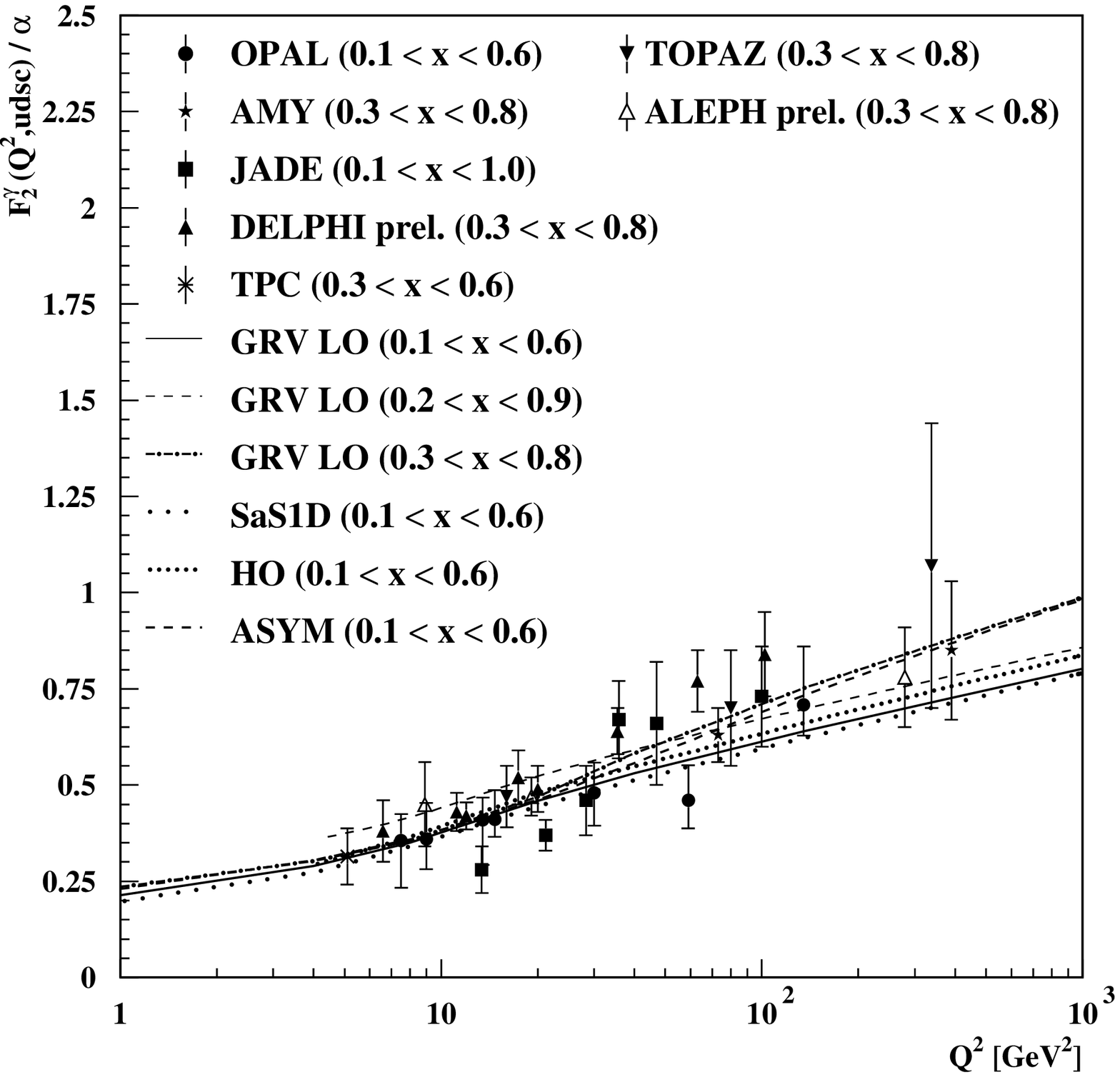,width=12.cm}
 \caption{%
 \it{
 Summary of the measurements of \ft\ at medium $x$ 
 without subtraction of the charm contribution
 compared to several theoretical predictions taking
 into account the various ranges in $x$.
 \label{fig:fig3}
 }
 }
 \end{center} \end{figure}

 The evolution of \ft\ with \qsq, Fig~\ref{fig:fig3},
 has been studied using the large lever
 arm in \qsq, and also by comparing various ranges in $x$ within one 
 experiment~\cite{OPALPR207}.
 Unfortunately the different experiments quote their results for different
 ranges in $x$ which makes the comparison more difficult because 
 the predictions for the various ranges in $x$ start to be significantly 
 different for $\qsq>100$~\gevsq, as can be seen in Fig~\ref{fig:fig3}.
 The measurements are consistent with each other and a clear rise of
 \ft\ with \qsq\ is observed.
 It is an interesting 
%
%
 fact that this rise can be described
 reasonably well (${\cal O}(15\%)$ accuracy) by
 the leading order asymptotic solution~\cite{WIT-7701}
 for \ft\ as predicted by perturbative QCD for
 $\almz=0.128$ as detailed in~\cite{OPALPR207}.
%
%
\section{Conclusion}
\label{sec:concl}
 New informations on the photon structure in so far unexplored 
 \qsq\ and $x$ regions have been obtained based on structure function 
 measurements at LEP.
 All observations are consistent with the QED and QCD predictions.
 The future LEP2 programme will allow to extend the study
 of the photon structure up to $\qsq\approx1000$~\gevsq.
 \par
 \vspace{0.2cm}
 {\bf Acknowledgement:}
 I wish to thank the members of the LEP collaborations for their 
 help during the preparation of this review.
 \\
 \vspace{-0.6cm}
%
%

%

\begin{thebibliography}{10}
 \bibitem{PLU-8102}
     PLUTO Collab., C.~Berger et~al.,
     Phys. Lett. {\bf 107B}, 168--172 (1981).
 \bibitem{ALE-EGM01}
     ALEPH Collab., contributed paper to Photon 97, Egmond aan Zee.
 \bibitem{L3C-JER01}
     L3 Collab., contributed paper to this conference.
 \bibitem{OPALPR182}
     OPAL Collab., K.~Ackerstaff et~al.,
     Z. Phys. {\bf C74}, 49--55 (1997).
 \bibitem{OPALPR088}
     OPAL Collab., R.~Akers et~al.,
     Z. Phys. {\bf C60}, 593--600 (1993).
 \bibitem{DEL-9601}
     DELPHI Collab., P.~Abreu et~al.,
     Z. Phys. {\bf C69}, 223--234 (1996).
 \bibitem{TPC-8401}
     TPC/2$\gamma$ Collab., M.~P. Cain et~al.,
     Phys. Lett. {\bf 147B}, 232--236 (1984).
 \bibitem{CEL-8301}
     CELLO Collab., H.~J. Behrend et~al.,
     Phys. Lett. {\bf 126B}, 384--390 (1983).
 \bibitem{PLU-8501}
     PLUTO Collab., C.~Berger et~al.,
     Z. Phys. {\bf C27}, 249--256 (1985).
 \bibitem{OPALPR185}
     OPAL Collab., K.~Ackerstaff et~al.,
     Z. Phys. {\bf C74}, 33--48 (1997).
 \bibitem{OPALPR207}
     OPAL Collab., K.~Ackerstaff et~al.,
     Phys. Lett. {\bf B411}, 387--401 (1997).
 \bibitem{OPALPR213}
     OPAL Collab., K.~Ackerstaff et~al.,
     Phys. Lett. {\bf B412}, 225--234 (1997).
 \bibitem{TIA-9701}
     DELPHI Collab., {\em Proceedings of Photon 97}, Egmond aan Zee.
 \bibitem{ALE-JER01}
     ALEPH Collab., contributed paper to this conference.
 \bibitem{GLU-9201}
     M.~Gl{\"u}ck, E.~Reya, and A.~Vogt,
     Phys. Rev. {\bf D45}, 3986--3994 (1992).
 \bibitem{GLU-9202}
     M.~Gl{\"u}ck, E.~Reya, and A.~Vogt, 
     Phys. Rev. {\bf D46}, 1973--1979 (1992).
 \bibitem{SCH-9501}
     G.~A. Schuler and T.~Sj{\"o}strand, 
     Z. Phys. {\bf C68}, 607--623 (1995).
 \bibitem{PLU-8401}
     PLUTO Collab., C.~Berger et~al.,
     Phys. Lett. {\bf 142B}, 111--118 (1984).
 \bibitem{TPC-8701}
     TPC/2$\gamma$ Collab., H.~Aihara et~al.,
     Z. Phys. {\bf C34}, 1--13 (1987).
 \bibitem{WIT-7701}
     E.~Witten, Nucl. Phys. {\bf B120}, 189--202 (1977).
\end{thebibliography}
\end{document}